\documentstyle[prd,epsfig,aps]{revtex}

\def\Mp{M_{\ast}}
\def\beqra{\begin{eqnarray}}
\def\eeqra{\end{eqnarray}}
\def\beq{\begin{equation}}
\def\eeq{\end{equation}}
\def\ds{\displaystyle}

\def\vp{\varphi}

\begin{document}
\twocolumn

\draft
\input epsf
\twocolumn[\hsize\textwidth\columnwidth\hsize\csname
@twocolumnfalse\endcsname

\title{Dynamical Relaxation of the Dark Matter to Baryon Ratio}
\author{R.~Catena$^{(1)}$,  M.~Pietroni$^{(2)}$ and L.~Scarabello$^{(2,3)}$} 
\address{$^{(1)}${\it Scuola Normale Superiore and INFN - Sezione di Pisa, 
Piazza dei Cavalieri 7, I-56125 Pisa, Italy  }}
\address{$^{(2)}${\it INFN, Sezione di Padova, via Marzolo 8, I-35131, Padova, Italy}}
\address{$^{(3)}${\it Dipartimento di Fisica, Universit\`a di Padova,  via Marzolo 8, I-35131, Padova, Italy }}

\address{\rm e--mail: r.catena@sns.it, pietroni@pd.infn.it, scarabel@pd.infn.it}

\maketitle
\begin{abstract}
A scalar field interacting differently with dark matter and baryons may explain why their ratio is of order unity today. We provide three working examples, checking them against the observations of CMB, Large Scale Structure, supernovae Ia, and post-newtonian tests of gravity. Such a scenario could make life much easier for supersymmetric dark matter candidates.
\end{abstract}

]
%
%
\section{Introduction}
%
One of the most debated puzzles in cosmology is why the ratio between the energy densities of Dark Energy (DE) and Dark Matter (DM) is around two today. Much less perceived as a disturbing `coincidence' is the fact that the ratio between the DM and baryonic energy densities is around five. 

From an aesthetic  point of view, the latter coincidence looks more digestible than the former. Indeed, contrary to the DE/DM case, the DM to baryon ratio is constant in time and its actual value of order unity is not a peculiarity of the present epoch. Still, there is a priori no reason why two uncorrelated physical processes, like baryogenesis and the generation of DM abundance, yield relics of roughly the same energy densities.

The purpose of this letter is to present a very simple scheme to address this issue: the ratio of the DM to baryon energy density is given by a ratio of coupling constants, like in eq.~(\ref{fixratio}), and so may naturally be of order unity. 

In our framework, baryons and DM couple differently to the same scalar field, so that the ratio between their energy densities, $\rho_b/\rho_d$, becomes a field-dependent quantity. Under general circumstances, the field equations of motion exhibit fixed points so that the ratio is driven from its early value, determined by baryogenesis and DM freeze-out, to a fixed point. 

The idea of a scalar field with different couplings to the dark and the visible sectors has a long history, dating back to the work of Damour and Gundlach \cite{damgun}, see also \cite{belqui}. 
More recently, dark matter particles with a time-dependent mass, were proposed as a way to increase the age of the universe \cite{andcar}, and, after the observation of the acceleration of the universe \cite{SNe}, as a way to solve the coincidence problem between DE and DM \cite{coinc}. More investigations of new scalar interactions in the dark sector were presented in \cite{farpeeb,gubpeeb}.

Our scenario differs from the previous ones mainly in two respects. First, the coupling of the scalar field to baryons was generally discarded in the past, both because the focus was mainly on the DE/DM sector and because it is a potentially dangerous source of fifth forces and violations of the equivalence principle. This coupling is, on the other hand, crucial here, in order to fix the DM to baryon ratio dynamically. 
Second, in models in which DM tracks DE, the growth of cosmological perturbations is deeply modified in the past, as a consequence of the new scalar interaction and of the exotic dilution rate for the DM energy density $\rho_{DM} \neq a^{-3}$  \cite{coinc}. 
Here, as we will see, the scalar field is fixed in the past and the growth of perturbation is just like in the standard $\Lambda$CDM case -- with at most small deviations in the recent past-- regardless of the strength and non-universality of the scalar interaction with DM and baryons.

Although the main features of our scenario  can be implemented in a generic model, in order to be definite we work in the framework of Scalar-Tensor theories of gravity with purely metric couplings but different conformal factors in the visible and the dark sectors, a generalization of the model studied in \cite{damgun}.

From a phenomenological point of view, Scalar-Tensor theories have the virtue of preserving the equivalence principle by construction, {\it i.e.} without the need to tune the coupling to the visible sector to extremely tiny values. From a theoretical point of view they are  an attractive way to accommodate ultralight scalar fields. Indeed, general covariance implies that the contribution of radiative corrections from the (visible and dark) matter sector to the scalar field mass is at most of order $\Lambda^4/M_p^2$, $\Lambda$ being the cosmological constant (see, for instance, \cite{choi}). Thus, the lightness of the scalar field is just a manifestation of the smallness of the cosmological constant, or of the curvature of the universe.

In the following, we will present the general framework (sect. II), and then give three possible realizations, depending on the form of the scalar effective potential (sects. III, IV and V). Then, in sect. VI, we will make some comments on the possible phenomenological impact of this scenario.

%
%
\section{Visible and dark sector}
%
We will consider an action of the form
\begin{equation}
S=S_{g}+S_{b}+S_{d}\,,  \label{action}
\end{equation}
where $S_g$ is the gravitational part, given by the sum of  the Einstein-Hilbert action and the scalar field one,
\begin{equation}
S_{g}=\frac{M_{\ast}^2}{2}\int d^{4}x\sqrt{-{g}}\left[ {R}+{g}^{\mu
\nu }\partial _{\mu }\varphi \partial _{\nu }\varphi -\frac{2}{M_{\ast}^2} V(\varphi )\right] ,
\end{equation}
whereas $S_b$ and $S_g$ represent the visible (or `baryonic') action and the dark one, respectively,
\begin{equation} 
S_{b,d} = S_{b,d}[\Psi_{b,d},A_{b,d}^{2}(\varphi ){g}_{\mu \nu }] \,\,\, , 
\label{sbd}
\end{equation}
with $\Psi_{b}$ ($\Psi_d$) indicating a generic field of the visible (dark) sector coupled to the metric
$A_{b}^{2}(\varphi ){g}_{\mu \nu }$ ($A_{d}^{2}(\varphi ){g}_{\mu \nu }$). Since, for our purposes, the most relevant component of the visible sector is the baryonic one, we will label fields, densities, etc., from this sector with the letter  `b'. $V(\vp)$ can be either a true potential or a (Einstein frame) cosmological constant, $V(\vp)=V_0$.

The emergence of such a structure, with different conformal factors for the visible and the dark sectors can be motivated in extra-dimensional models, assuming that the two sectors live in different portions of the extra-dimensional space. One could also generalize the above structure to include more dark matter species, each with a different conformal factor, as was done in \cite{gubpeeb}. Moreover, it should be noticed that our results hold for a more general lagrangian than the one above. In practice, as we will see, all we need is that the ratio of baryon masses to dark matter ones has a non-trivial field dependence. In order to preserve the equivalence principle, we have to impose that the ratio of proton to neutron masses is constant, which is the case for the action in eq.~(\ref{sbd}).

If $A_b$ is constant, then  the $S_g + S_b$ sector is just that  of GR, plus a scalar field decoupled from the physical sector.  The {\it distance} to GR is then measured by the quantity
 \beq 
 \alpha_b = \frac{d \log A_b}{d \vp}\,.
\label{alphab}
\eeq

We consider an homogeneous cosmological space-time 
\[
ds^{2}= dt^{2} - a^{2}(t)\ dl^{2}\ , 
\]
where the matter energy-momentum tensors, $T^{b,d}_{ \mu \nu } = 
2(-g)^{-1/2}   \delta S_{b, d} / \delta g^{\mu\nu}$  admit the perfect-fluid representation 
\[T^{b,d}_{ \mu \nu } 
=(\rho_{b,d} +p_{b,d})\ u_{\mu }u_{\nu }\ - p_{b,d}\ g_{\mu \nu }\ \ , 
\]
with $g_{\mu \nu }\ u^{\mu }u^{\nu }=1$.

The Friedmann-Robertson-Walker (FRW) equations then take the form 
\begin{eqnarray}
&& \frac{\ddot{a}}{a} = - \frac{1}{6 \Mp^2} \left[ \rho_b +3\ p_b +\rho_d +3\ p_d +2 \Mp^2\dot\varphi^{2}-2V \right]\nonumber\\
&&\\
&&\nonumber\\
&&\quad\left( \frac{\dot{a}}{a}\right) ^{2}+\frac{k}{a^{2}} = \frac{1}{3 \Mp^2}
\left[\rho_b+\rho_d + \frac{\Mp^2}{2}\dot\varphi^{2}+V \right] \nonumber\\
&&\\
&&\nonumber\\
&&\ddot{\varphi} +3\frac{\dot{a}}{a}\dot\varphi = 
-\frac{1}{M_{\ast}^2}\left[ \alpha_b (\rho_b-3p_b)+ \alpha_d (\rho_d-3p_d)+\frac{\partial V}{\partial \varphi } 
 \right]\nonumber\\
 &&
 \label{eqback}
\end{eqnarray}
where $\alpha_d$ is the analogous of eq.~(\ref{alphab}), and  the Bianchi identity is
\begin{equation}
{}d(\rho_{b,d} \,a^{3})+p_{b,d}\ da^{3}=(\rho_{b,d} -3\ p_{b,d})\ a^{3}d\log A_{b,d}(\varphi ). 
\label{bianchi}
\end{equation}
In the following, we will consider only  pressureless matter both in the visible and in the dark sector, which by eq.(\ref{bianchi}) scale as
\beq
\rho_{b,d} \sim A_{b,d}(\vp) a^{-3}\,.
\label{scale}
\eeq
The main point of this paper is that if $A_b(\vp) \neq A_d(\vp)$ the ratio $\rho_b/\rho_d$ becomes a dynamical quantity with a  non-trivial $\vp$-dependence. Then the present day value of this ratio is determined, in principle, both by the early-time mechanisms of baryogenesis and freeze-out and by the late time behavior of the field $\vp$. 
 
The above equations have been gven in the so-called `Einstein frame'. Actually, an observer using non-gravitational `visible' units, like atomic clocks, wavelenghts, and masses,   is using `physical (or Jordan) frame' units, defined by the rescaled metric
\beq\tilde{g}_{\mu \nu } = A^2_b(\vp) g_{\mu\nu}\,.
\eeq
In this frame, $\tilde{\rho}_b$ has the canonical scaling law, $\tilde{\rho}_b \sim \tilde{a}^{-3}$, whereas 
$\tilde{\rho}_d \sim A_{d}(\vp) A_{b}(\vp)^{-1}  \tilde{a}^{-3}$, the ratio being frame-independent. $\tilde{a}=a A(\vp)$ is the scale factor in the visible frame.
In the following, we will solve the equations of motion in the Einstein frame, which is more practical, and will then translate the results in the physical one.

Next, we consider the linear perturbations. We use as basis variables the perturbation of the scalar field, $\delta \vp$, and  the combination,
\beq
\delta_{d,b } \equiv \delta \rho_{d,b}/\rho_{d,b} - \alpha_{d,b} \, \delta \vp\,,
\eeq
which, since
\beq
\frac{\rho_{d,b}}{A_{d,b}} \sim a^{-3} \sim n_{d,b}\,,
\eeq
 can be identified with the linear perturbation in the particle number density. The use of  $\delta_{d,b }$ instead of $\delta \rho_{d,b}/\rho_{d,b} $ is motivated by the fact that the former  are $\delta \vp$-independent, as we can read from eq.~(\ref{scale}), so that we can single out $\delta \vp$ and compute its effective mass properly (see eq.~(\ref{effmass})). Notice, however, that energy and number density perturbations tend to coincide in the Newtonian limit,  $k^2/ a^2 H^2 \gg 1$ ($k$ being the comoving momentum), in which they  differ only by $O(\alpha^2_{b,d} a^2 H^2/k^2)$ terms. Indeed, in this limit, the equation for the scalar field perturbation $\delta \vp$ gives
\beq 
\delta \vp \simeq -3 \frac {a^2 H^2}{k^2} \left(1+\frac{a^2 m_\vp^2}{k^2}\right)^{-1} (\alpha_d \Omega_d \delta_d + \alpha_b \Omega_b \delta_b)\,.
\eeq

The effective  mass $m_\vp$ has two contributions,
\[m_\vp^2 = m^2_V + m^2_\rho ,\]
with,
\beq
m^2_V = \frac{1}{\Mp^2}\frac{d^2 V}{d \vp^2}\,,\;\;\;\;\; m^2_\rho = \sum_{i=d,b} \frac{\rho_i}{\Mp^2} \left( \alpha_i^2+ \frac{d \alpha_i}{d \vp}\right) \,.
\label{effmass}
\eeq
In ref.~\cite{Amendola} the first term in brackets in $m_\rho^2$ was missed as a consequence of discussing perturbations in the $(\delta \rho_{d,b}/\rho_{d,b} ,\,\delta \vp)$ basis.

In the Newtonian limit, number  density perturbations satisfy the the same equations as the energy density perturbations, that is,
\beqra
&& \delta_d^{\prime \prime} +\left(2+\frac{H^\prime}{H} +\alpha_d \vp^\prime\right) \delta_d^{\prime}  - \frac{3}{2} (\gamma_{dd} \Omega_d \delta_d + \gamma_{db} \Omega_b \delta_b)=0\,,\nonumber\\
&& \nonumber\\
&& \delta_b^{\prime \prime} +\left(2+\frac{H^\prime}{H} +\alpha_b \vp^\prime\right)  \delta_b^{\prime} - \frac{3}{2} (\gamma_{bd} \Omega_d \delta_d + \gamma_{bb} \Omega_b \delta_b)=0\,,\nonumber\\
&&
\label{pertgro}
\eeqra
where 
\beq 
 \gamma_{ij} = 1 + 2 \alpha_i \alpha_j \left(1+\frac{a^2 m_\vp^2}{k^2}\right)^{-1} \,\,\,\,(i,j = b, d)\,,
 \label{gammaij}
 \eeq
from which one can clearly read the extra force due to the scalar field, giving a non-universal coupling effective at scales such that $k^2 > a^2 m_\vp^2$. This gives a species-dependent Yukawa correction to the force between two trace
pointlike and  static masses as
\beq
F_{ij} = \frac{G m_i M_j}{r^2} \left( 1+ 2 \alpha_i \alpha_j e^{-m_\vp r}\right)\,,
\label{Newton}
\eeq 
where $(i,j = b,d)$.
As we will discuss in sect.~\ref{cam}, if the masses are not pointike, the  couplings 
appearing in (\ref{Newton}) can be effectively much smaller than the `bare' values
$\alpha_{i,j}$.

The corresponding equations in the physical frame can be obtained straightforwardly. Indeed, since $\tilde{\rho}_i= A_b^{-4} \rho_i$  $(i=d,\,b)$,  we have
\beq
\tilde{\delta}_{b,d} = \delta_{b,d} - 4 \alpha_b \delta \vp \,,
\eeq
so that the two coincide in the Newtonian limit. Moreover, since $\tilde{m}_\vp \tilde{a} /k= m_\vp a/k$, the translation from the Einstein equations above to the physical frame simply amounts to the change of variable
\beq
d \log \tilde{a} = d \log a + \alpha_b d\vp \,.
\eeq

In the following, we will discuss three explicit realizations of the dynamical determination of the baryon to dark matter ratio, considering three different possibilities for the scalar potential $V(\vp)$.

\section{The cosmological constant}
As a first example, we consider the case in which the Einstein frame potential is just a constant. Then, if both visible and dark matter are pressureless,  the right hand side of eq.~(\ref{eqback}) is proportional to the function,
\beq
 F(\vp)\,a^{-3} \equiv \alpha_b \rho_b + \alpha_d \rho_d\,,
 \eeq 
 where the $\vp$-dependence resides both in the $\alpha's$ and in the $\rho's$, see eq.~(\ref{scale})), and $a_0 =1$.

The zeroes of the function $F(\vp)$ determine the behavior of the system.
If $F(\vp)$ has no zeroes, then the field $\vp$ runs forever and the $\rho_b/\rho_d$ ratio is therefore time dependent. The Newton constant in the physical frame is also time dependent if $\alpha_b\neq0$, since $\dot{G}/G \sim \tilde{H} \alpha_b d \vp/d\log a $ and, as discussed in refs.~\cite{coinc,farpeeb,amquer}, the non-standard expansion rate and non- growth of perturbations may run into problem with CMB and large scale structure observations.

We will concentrate on the case in which  $F(\vp)$ has zeroes. Then, it is easy to realize that  a zero with a positive (negative) slope  is an attractive (repulsive) fixed point for the field evolution. Depending on the initial conditions, the field will proceed towards the next  zero with a positive slope, which will determine the late-time cosmology. In particular, the ratio of densities will be given by
 \beq
  \frac{\rho_b}{\rho_d} = - \frac{\alpha_d(\vp_0)}{\alpha_b(\vp_0)}\,,
 \label{fixratio}
 \eeq
 $\vp_0$ being the attractive zero of $F(\vp)$ mentioned above. The dynamics of the approach to the fixed point depends on the form of the function $F$. In general, the steeper $F$ around $\vp_0$, the faster the field oscillations around it are damped. We will discuss this point more explicitly in sect.~\ref{dynde}. 
 On the fixed point the perturbaions exhibit an interesting feature. We look for solutions of the equations~(\ref{pertgro}) of the form $\delta_d \sim a^m$, $\delta_b= b \delta_d$. Due to the dependence of the $\gamma_{ij}$ on the $\alpha$'s and on $k$, the exponent $m$ and the bias parameter $b$ are generically expected to be $\alpha_{b, d}$ and $k$-dependent too. However, due to the relation in (\ref{fixratio}), a `miracle' happens. We find the solutions
\beqra
&&  b=1\,\,\,\,\,\, {\mathrm and} \,\,m=1, -3/2\,,\nonumber \\
&& \nonumber \\
&&  \ds b=\frac{\alpha_b}{\alpha_d}\,\,\,\,\,\, {\mathrm and} \,\,m=\frac{-1\mp \sqrt{1-48 \alpha_b \alpha_d 
(1+a^2 m_\vp^2/k^2)^{-1}}}{4}\,,\nonumber\\
\label{modes}
 \eeqra
 where we recognize the ($k$-independent) $m=1$, $b=1$ mode typical of a standard matter dominated epoch. Thus, independently on the strengths of the scalar interactions, the stationarity condition (\ref{fixratio}) implies that the fluctuations of the two fluids will always have the standard adiabatic growing mode.  The other mode corresponds to a potentially dangerous isocurvature perturbation ($\Omega_d \delta_d +\Omega_b \delta_b=0$).
However, the growth of this mode is extremely suppressed once the post-newtonian bounds are considered, as we will discuss now. 
 
The effective mass on the fixed point is
\beq
m^2_\vp = m^2_\rho=\frac{\rho_d \alpha_d}{\Mp^2} \left(\alpha_d-\alpha_b+ \frac{d \log(\alpha_d/\alpha_b)}{d \vp}\right)\,,
\eeq
which is $O(\alpha_{b,d} H_0^2)$ . Then, unless one of the couplings $\alpha_{b,d}$ is enormous, the scalar field is massless on solar system scales and the coupling to visible matter is bounded by solar system tests of gravity. The recent analysis of the delay of radio signals from the Cassini spacecraft \cite{Cassini} gives the stringent upper bound 
\beq 
\alpha^2_b  <  2.4 \times 10^{-5} \,\,({\mathrm at}  \, 3\, \sigma).
\label{cassini}
\eeq
$\alpha_d$ is then fixed by eq.~(\ref{fixratio}), once the best fit value for the $\rho_b/\rho_d$ ratio \cite{cosmparam},
\beq
\frac{\rho_b}{\rho_d} = 0.20 \pm 0.02\,,
\eeq has been imposed. 
Other astrophysical and cosmological bounds on $\alpha_d$, as discussed for instance by Frieman and Gradwhol in \cite{belqui}, would be much less restrictive. 

As anticipated, we conclude that the isocurvature mode is likely to play no role, since the bounds above, once inserted in eq.~(\ref{modes}), imply that the growth exponent $m$  is at most $O(10^{-3})$, to be compared to unity for the adiabatic one.

Thus, once the field has settled to the fixed point, the cosmology is just the standard one apart from the value of the $\rho_b/\rho_d$ ratio which would be fixed by $\vp_0$ rather than by the standard mechanism for the generation of the relic abundance via freeze-out. 

\section{The  Chameleon case}
\label{cam}
Now we consider the case of a quartic potential,
\beq
V=V_0 + \bar{V}  \,\frac{(\vp - \vp_0)^4}{4 !}.
\eeq
Indeed, as we will see shortly, a potential of this type can fix the $\rho_b/\rho_d$ ratio, but cannot account for DE, so that a cosmological constant term $V_0$ should be added.
The field attractive fixed point is now given by
\beq 
-(\vp - \vp_0) = \left( 6 \, \frac{\alpha_b \rho_b+\alpha_d \rho_d}{\bar{V}}\right)^{1/3}\,,
\eeq
where it is assumed that $ (\alpha_b \rho_{b}+\alpha_d \rho_d)/\bar{V} \ll 1$.  $\vp$ is then squeezed very close to the minimum of the potential, which fixes the $\rho_b/\rho_d$ ratio. The contribution of the field-dependent part of the potential to the total energy density is suppressed by a large $\bar{V}$,
\beq
\frac{V(\vp) - V_0}{\rho_b+\rho_d} = \left( \frac{\rho_b+\rho_d}{\bar{V}}\right)^{1/3} \, \left(6 \frac{\alpha_b \rho_b+\alpha_d \rho_d}{\rho_b+\rho_d}\right)^{4/3}\ll 1\,,
\eeq
and, as we anticipated, it cannot account for the observed DE if $\rho_{b,d}\ll \bar{V}$.

By the same mechanism, the field evolution takes place on a timescale much larger than the Hubble one, so that we have, for instance
\beq
\left|\frac{\dot{G}}{G}\right| \sim \tilde{H} \left|\alpha_b \frac{d\vp}{d\log a}\right|\sim \tilde{H} \left|\alpha_b\right| \left( 6 \, \frac{\alpha_b \rho_b+\alpha_d \rho_d}{\bar{V}}\right)^{1/3} \ll \tilde{H}\,,
\eeq
which can be safely reconciled with existing bounds on the variation of $G$.

The scalar field gets an effective mass
\beqra
&& m_\vp^2 \simeq m_V^2=\frac{1}{\Mp^2}\frac{d^2 V}{d \vp^2} \nonumber\\
&&\quad\quad = \frac{\rho_b+\rho_d}{2 \Mp^2} \left(\frac{\bar{V}}{\rho_b+\rho_d}\right)^{1/3} \left( 6 \, \frac{\alpha_b \rho_b+\alpha_d \rho_d}{\rho_b+\rho_d}\right)^{2/3},
\eeqra
which, using the value of the cosmological density of matter today, gives a range of order
\beq
m_\vp^{-1} \sim 5 \cdot 10^2 \,{\mathrm Km} \left(\frac{\alpha_d^2 \bar{V}}{\Mp^4}\right)^{-1/6}\,.
\label{massa}
\eeq
For $\alpha_d =O(1)$ or larger, the linear perturbations would be affected by the scalar field only at scales smaller than about 100 Km. For smaller $\alpha_d$ the scalar field range increases, but at the same time the coupling decreases, so that no dramatic effect on the growth of perturbations is expected, at least at linear order. 

Since the mass in eq.~(\ref{massa}) is much larger than $H$ the approach to the minimum of the chameleon potential takes place on a time-scale typically much smaller than the age of the universe.

Inside our Galaxy ($\rho \sim 10^{-24} {\mathrm g/cm^3}$) the range is about fifty times smaller than eq.~(\ref{massa}). At first sight, such macroscopic ranges appear to be in conflict with fifth force experiments in the lab or in the solar system. However, as it was emphasized in \cite{chameleon}  the density-dependence of the effective mass greatly reduces the strength of the existing bounds. Indeed, inside baryonic test bodies the effective mass is of order
\beq
m_{\vp\, \mathrm{in}}^{-1} \sim 0.1\, {\mathrm mm} (\rho_b [\mathrm{g/cm^3}])^{-1/3} \left(\frac{\alpha_b^2 \bar{V}}{\Mp^4}\right)^{-1/6}\,.
\eeq
If the size $R$ of the body is much larger than $1/ m_{\vp\, \mathrm{in}}$ the field profile is flat inside it except for a shell of size $\sim 1/m_{\vp\, \mathrm{in}}$ and the scalar force do not penetrate inside it, a phenomenon analogous to the Meissner effect for superconductors, which was dubbed `thin-shell effect' in ref. \cite{chameleon}. In this context, it can be described as an effective coupling
\beq
\alpha_b^{eff} = \left(\frac{3}{m_{\vp\, \mathrm{in}} R}\right)^2 \alpha_b\,,
\eeq
replacing $\alpha_b$ in eq.~(\ref{Newton}) if $m_{\vp\, \mathrm{in}} R\gg 1$.  Following ref. \cite{chameleon}, the existing tests  on a fifth force translate in an upper bound of order
\beq
\alpha_b \frac{\Mp^4}{\bar{V}} < O(10)\,, 
\eeq
relaxing the bound on  $\alpha_b$ by many orders of magnitude with respect to eq.~(\ref{cassini}). 
Geophysical and astrophysical searches of deviations from General Relativity give rise to much  weaker bounds, due to the larger amount of suppression caused by the thin-shell effect in the Earth, the Moon, and the Sun.

\section{Dynamical DE}
\label{dynde}
In the previous two sections the role of DE was played by a cosmological constant. Here, we examine the possibility that the same field $\vp$ controlling the $\rho_b/\rho_d$ ratio also accounts for a time-dependent DE. In turn, this will imply  time-dependent $\rho_b/\rho_d$ ratio, $\alpha_b$, and $\alpha_d$,  though only for the very last red-shifts. All this might lead to conflicts with cosmological observations. 

Indeed, the expansion rate in the physical frame is given by,
\beq
\tilde{H} = \frac{\dot{a}}{a} A_b(\vp)^{-1} \left(1+\alpha_b \frac{d \vp}{d \log a}\right) .
\eeq
We must check that it differs by no more than 10 \% from the standard one at Big Bang Nucleosynthesis \cite{nubound}, that it doesn't shift the CMB peaks by more than 3 \%, as measured by the `shift parameter' \cite{shift}
\beq
{\cal R}_0 = \Omega_m^{1/2} \tilde{H}_0 \, r(z_{dec}) = 1.716 \pm 0.062 ,
\eeq
$r(z_{dec})$ being the distance to the decoupling surface,
and that it gives an acceptable age for the universe, $\tilde{H}_0 \tilde{t}_0 = 0.97\pm 0.01$, where we have used the results from WMAP at $68 \%$ c.l. \cite{wmap}.

At small redshifts, the growth of perturbations will depart from the standard solution of a matter-dominated epoch, $m=b=1$, not only because DE is overtaking DM, but also because, when the field moves,  the stationarity condition of eq.~(\ref{fixratio}) is violated. We will then use the constraint from 2dF survey on $\beta= f/b= 0.49 \pm 0.09$  \cite{2df} and $b=1.04\pm 0.11$ \cite{verde} at the effective redshift 0.15, to get
\beq
f(\tilde{z}\sim 0.15)=0.51\pm 0.11,
\eeq
where $f=d \log \delta_d/d\log \tilde{a}$.

Moreover, we must check the resulting expansion rate at recent epochs against the supernovae data \cite{SNe}. We tested the distance-redshift relations of our models against the ``gold" set of 157 SNe Ia of ref. \cite{sngold}, using flux-averaging statystics as implemented in \cite{wang}.

On top of that, the post-newtonian bound of eq.~(\ref{cassini}) must be satisfied at $\tilde{z}=0$.

In order to be be definite, we must chose a functional form for the scalar potential and the functions $A_{b,d}(\vp)$. Of course, lacking a deep theoretical inspiration, a degree of arbtrariness is present at this stage. However, since -- after the early-time relaxation to the fixed point -- the field starts to move again only at small redshifts, its late-time behavior mostly depends on the values of the functions and their lower derivatives rather than on the full functional form. Therefore, the main features of the results we will present are independent on the particular functional choice.

 $A_{b}(\vp)$ and $A_{d}(\vp)$ must have opposite slopes, in order that their logarithmic derivatives satisfy the condition (\ref{fixratio}). Moreover, $A_b$ should be extremely flat today in order to meet the post-newtonian bounds of eq.~(\ref{cassini}). A natural way to implement this, is to assume a $Z_2$ symmetry, $\vp \rightarrow -\vp$, which sets to zero the first derivatives of $A_b$ and $A_d$ at the fixed point $\vp = 0$. Thus, taking into account that without loss of generality we can take $A_b=A_d=1$ on the fixed point, we can consider the expansions
 \beq
A_b = 1 + b_b \vp^2 + \cdots \,,\;\;\;\;A_d =  1 + b_d \vp^2 + \cdots\,,
\label{chA}
\eeq
and
\beq
V(\vp)=V_0 + V_2 \vp^2 +V_4 \vp^4 +\cdots.
\label{chV}
\eeq
Some illustrative results of solutions passing all the above mentioned tests are reported in the figures, together with the $\Lambda$CDM case (continuous line).

\begin{figure*}[t]
\epsfxsize=3.2 in 
\epsfysize=2.8 in
\epsfbox{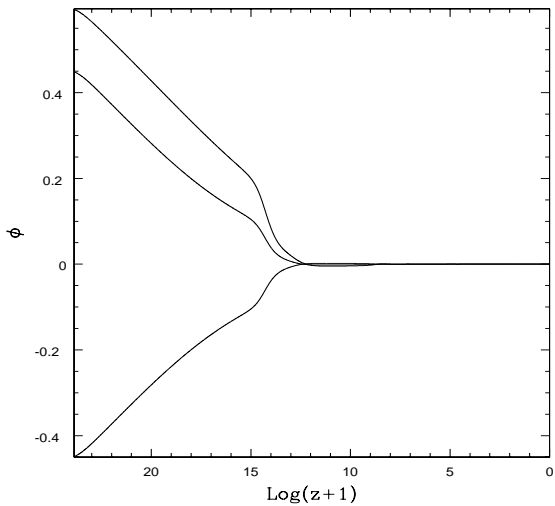}
\caption{The evolution of the field at early time with different initial conditions. The fixed point is reached well before nucleosynthesis. In this example we used $b_b=-0.05$, $b_d=5$. }
\label{plot_field}
\end{figure*}

In fig.~\ref{plot_field} we show the evolution of the scalar field from a very early epoch to the present one. As we see, the attraction to the fixed point is effective, so that it is reached well before nucleosynthesis. Notice that during radiation domination the drag towards the fixed point is mainly due to the massive particles in the thermal bath. Each time one of these becomes non relativistic as the temperature decreases, the field receives a `kick' towards the fixed point, as was discussed for instance in ref. \cite{catena}. Moreover, the trace anomaly gives also a contribution to the field evolution, which we have also included in this paper.

\begin{figure*}[t]
\epsfxsize=3.2 in 
\epsfysize=2.8 in
\epsfbox{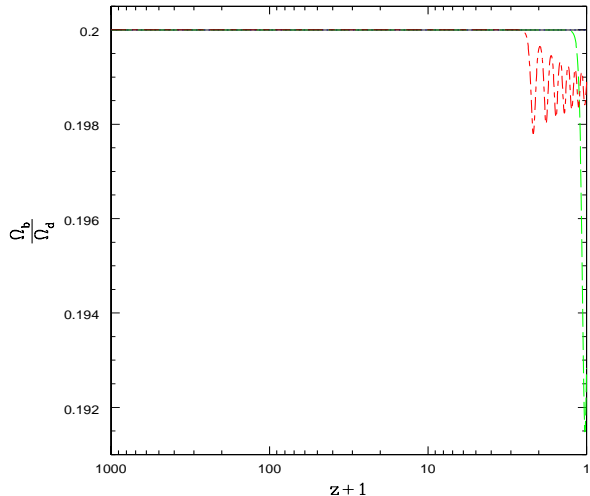}
\caption{The ratio between baryons and dark matter energy densities for three illustrative
examples satisfying all the experimental bounds discussed in the text. The solid line corresponds to the $\Lambda$CDM model with $\Omega_\Lambda = 0.7$.
The  dotted ( $V_2=-65 V_0$, $V_4=V_0$, $b_b=-0.05$, $b_d=5$),
dashed ( $V_2=-100 V_0$, $V_4=10^4 V_0$, $b_b=-0.01$, $b_d=5$) and  dot-dashed 
( $V_2=-500 V_0$, $V_4=2 \cdot 10^5 V_0$, $b_b=-0.05$, $b_d=5$) lines correspond to models passing all the tests mentioned in the text.
In each case  $V_0$ is adjusted to give the right amount of DE today.}
\label{plot_omega_2}
\end{figure*}

As a consequence of the field evolution the $\rho_b/\rho_d$-ratio evolves from its early-time value, set by baryogenesis and WIMP decoupling, to the fixed point value of eq.~(\ref{fixratio}). Afterwards, it remains practically constant through the epochs of nucleosynthesis and matter-radiation decoupling. Finally, depending on the sign of $V_2$, it may or may not start a late-time evolution, see fig.~\ref{plot_omega_2}.  

\begin{figure*}[t]
\epsfxsize=3.2 in 
\epsfysize=2.8 in
\epsfbox{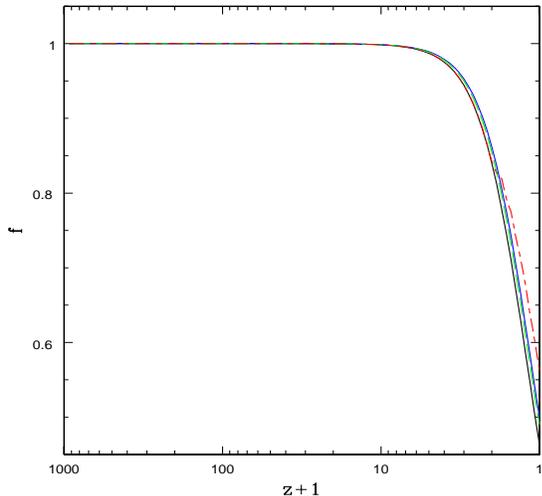}
\caption{Evolution of the growth exponent of the linear perturbations, f.}
\label{plot_f}
\end{figure*}

A positive $V_2$ gives a behavior practically indistinguishable from a pure cosmological constant. On the other hand, if $V_2$ is negative, a $Z_2$-breaking phase transition takes place when $\alpha_b\rho_b+\alpha_d \rho_d+dV/d\vp$ becomes negative, which happens at small redshifts. The scalar field then rolls down from the maximum at $\vp=0$ and  the $\rho_b/\rho_d$ ratio changes. The sign of change can be reversed by reversing the signs of $b_b$ and $b_d$. The smallness of the effect is due mainly to the post-newtonian bound of eq.~(\ref{cassini}) and to the short time available for the field to run. 
For the same reason, the perturbation growth (fig.~\ref{plot_f}) also follows a pattern quite close to $\Lambda$CDM, with $f$ differing at most by 10\% at redshift less than one.

\begin{figure*}[t]
\epsfxsize=3.2 in 
\epsfysize=2.8 in
\epsfbox{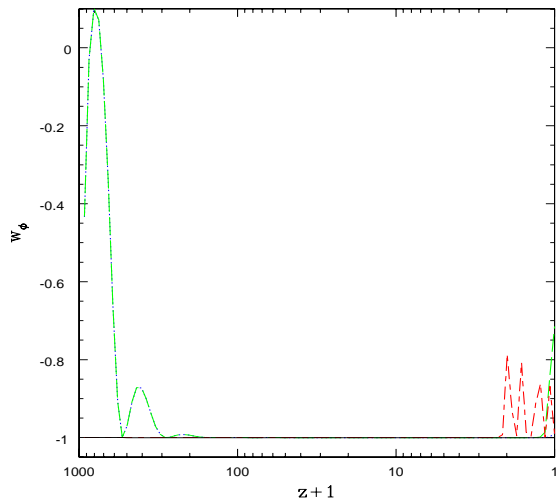}
\caption{Equation of state for the scalar field.}
\label{plot_W}
\end{figure*}

The equation of state (fig.~\ref{plot_W}) of the scalar field could be as high as $W=-0.7$ and still pass the SNe test at 95 \% c.l.. 

\begin{figure*}[t]
\epsfxsize=3.2 in 
\epsfysize=2.8 in
\epsfbox{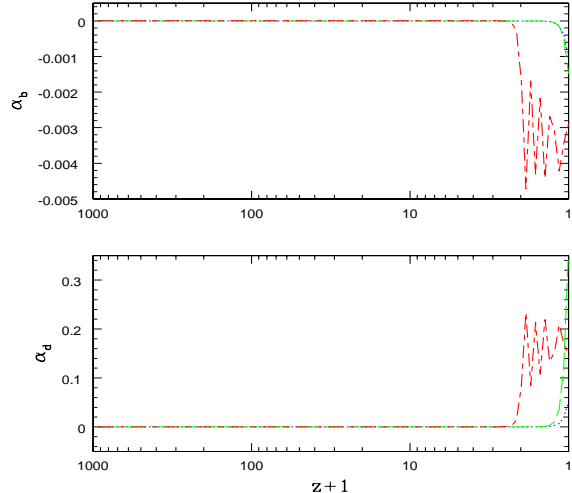}
\caption{The coupling functions $\alpha_b$ and $\alpha_d$ in the 
examples of fig. 2}
\label{plot_BD}
\end{figure*}

\section{conclusions}
The most popular DM candidate is the lighest supersymmetric particle, which, being stable and weakly interacting, has all the good qualities to be a successful WIMP. 
However, in practice, experimental constraints restrict the allowed parameter space of the Minimal Supersymmetric Standard Model (MSSM) to regions where the neutralino relic density is typically one order of magnitude below the WMAP preferred range (see, for instance, ref.~\cite{profumo}). Bino-like neutralinos, as preferred in the mSugra subclass of MSSM, suffer from the opposite problem, being usually overproduced because of their lower annihilation rate. As a result, the allowed parameter space of MSSM after the WMAP results has shrunk to regions were extra annihilation channels are present, which requires an increasing degree of fine tuning. 

The scenario presented in this paper has the potential to alleviate this problem considerably. The DM relic abundance produced at neutralino freeze-out, can be `adjusted' by the late time dynamical mechanism presented here, thus rescuing wide portions of the parameter space. Of course, once the coupling functions are chosen, the fixed point attraction should be strong enough as to drive the initial DM abundance to the WMAP value before nucleosynthesis, which would provide the limits of the region of the rescuable parameter space.

Other post-freeze-out adjustment mechanisms were proposed in \cite{salati} and in \cite{catena}. In both cases the adjustment could only be an enhancement of the relic density, whereas here we could have a reduction as well, a case of interest for a bino-like neutralino.

An obvious signature of these adjustment mechanism would be a laboratory discovery of 
a DM particle whose interactions do not match the relic abundance computed in the standard way. 

On the cosmological/gravitational side, an improvement of the post-newtonian bounds and an increase in the sensitivity to the linear growth rate $f$ by at least a factor of two might constrain this scenario considerably.

\section{Acknowledgment}
We would like to thank M. Liguori for the friendly technical support.

\end{document}